\documentclass[showkeys,superscriptaddress,amsmath,amssymb,aps,prl,twocolumn]{revtex4-2}

\usepackage{graphicx}
\usepackage{dcolumn}
\usepackage{bm}
\usepackage{xcolor}
\usepackage{amsmath}
\usepackage{chngcntr}
\usepackage{nicefrac}
\usepackage{hyperref}

\newcommand{\beginsupplement}{%
        \newpage ~\\
        \newpage
        \setcounter{table}{0}
        \renewcommand{\thetable}{S\arabic{table}}%
        \setcounter{figure}{0}
        \renewcommand{\thefigure}{S\arabic{figure}}%
     }

\begin{document}

\title{Observation of orbital order in the Van der Waals material $1T$-TiSe$_2$}

\author{Yingying Peng}
\email{yingying.peng@pku.edu.cn}
\affiliation{International Center for Quantum Materials, School of Physics, Peking University, Beijing 100871, China}

\author{Xuefei Guo}
\affiliation{Department of Physics and Materials Research Laboratory, University of Illinois, Urbana, IL 61801, USA} 

\author{Qian Xiao}
\affiliation{International Center for Quantum Materials, School of Physics, Peking University, Beijing 100871, China}

\author{Qizhi Li}
\affiliation{International Center for Quantum Materials, School of Physics, Peking University, Beijing 100871, China}

\author{J\"{o}rg Strempfer}
\affiliation{Advanced Photon Source, Argonne National Laboratory, Argonne, Illinois 60439, USA}

\author{Yongseong Choi}
\affiliation{Advanced Photon Source, Argonne National Laboratory, Argonne, Illinois 60439, USA}

\author{Dong Yan}
\affiliation{School of Materials Science and Engineering, State Key Laboratory of Optoelectronic Materials and Technologies, Sun Yat-Sen University, Guangzhou 510275, China}
\affiliation{Key Laboratory of Functional Molecular Solids, Ministry of Education, College of Chemistry and Materials Science, Anhui Normal University, Wuhu 241002, China}

\author{Huixia Luo}
\affiliation{School of Materials Science and Engineering, State Key Laboratory of Optoelectronic Materials and Technologies, Sun Yat-Sen University, Guangzhou 510275, China}

\author{Yuqing Huang}
\affiliation{International Center for Quantum Materials, School of Physics, Peking University, Beijing 100871, China}

\author{Shuang Jia}
\affiliation{International Center for Quantum Materials, School of Physics, Peking University, Beijing 100871, China}

\author{Oleg Janson}
\affiliation{Institute for Theoretical Solid State Physics, IFW Dresden, Helmholtzstr. 20, 01069 Dresden, Germany}

\author{Peter Abbamonte}
\affiliation{Department of Physics and Materials Research Laboratory, University of Illinois, Urbana, IL 61801, USA} 

\author{Jeroen van den Brink}
\affiliation{Institute for Theoretical Solid State Physics, IFW Dresden, Helmholtzstr. 20, 01069 Dresden, Germany}
\affiliation{W\"urzburg-Dresden Cluster of Excellence ct.qmat, TU Dresden, 01069 Dresden, Germany}
\affiliation{Institute for Theoretical Physics Amsterdam, University of Amsterdam, Science Park904, 1098 XH Amsterdam, The Netherlands}

\author{Jasper van Wezel}
\email{vanwezel@uva.nl}
\affiliation{Institute for Theoretical Physics Amsterdam, University of Amsterdam, Science Park904, 1098 XH Amsterdam, The Netherlands}

\date{\today}

\begin{abstract}
Besides magnetic and charge order, regular arrangements of orbital occupation constitute a fundamental order parameter of condensed matter physics. Even though orbital order is difficult to identify directly in experiments, its presence was firmly established in a number of strongly correlated, three-dimensional Mott insulators. Here, reporting resonant X-ray scattering experiments on the layered Van der Waals compound $1T$-TiSe$_2$, we establish the emergence of orbital order in a weakly correlated, quasi-two-dimensional material. Our experimental scattering results are consistent with first-principles calculations that bring to the fore a generic mechanism of close interplay between charge redistribution, lattice displacements, and orbital order. It demonstrates the essential role that orbital degrees of freedom play in TiSe$_2$, and their importance throughout the family of correlated Van der Waals materials.
\end{abstract}

\maketitle 

\section{Introduction}
Quasi two-dimensional Van der Waals (VdW) materials are layered solids with
strong in-plane covalent bonding and weak interlayer VdW interactions, that
have become a focal area for materials research in recent
years~\cite{otrokov19, li19, chen19, macneill19, gong19, kang20, wu20,
devarakonda21, noguchi21, diego21}. The success of graphene -- which stems from
the paradigmatic VdW material graphite -- in particular, triggered a search for
similar VdW materials that can be exfoliated to the monolayer limit, but which
harbour physical properties beyond those of graphene~\cite{Geim13,Novoselov16}. Magnetic ordering for
example, has been observed in the VdW materials Cr$X_3$ ($X$ = Cl, Br,
I)~\cite{li19, chen19, macneill19}, Cr$_2$Ge$_2$Te$_6$~\cite{gong19} and
NiPS$_3$~\cite{kang20}. 

The family of transition-metal dichalcogenides (TMDC) is a particularly
promising group of VdW materials, that can be straightforwardly prepared in
atomically thin configurations and device settings~\cite{Novoselov16}. They typically harbour
multiple competing and coexisting phases of matter, which allow a fine tuning
of the phase diagram and material properties in response to external stimuli
like pressure, intercalation, or gating~\cite{Kusmartseva09,Morosan06,Li16}. At the same time, however, uncertainty
and controversy surround the ground state order in several TMDC. This is
especially striking in the case of $1T$-TiSe$_2$, whose charge order (CDW) below
$T_{\text{CDW}} \simeq 200$\,K was discovered decades ago~\cite{disalvo76}. Debates about its nature and driving mechanism, however, are only recently converging towards a combination of exciton formation and lattice effects stabilising a condensate of particle-hole pairs~\cite{Rossnagel02, Wilson77,Cercellier07,Hughes77, Suzuki85,Kidd02,Calandra11,Kogar17,vanwezel10}, while debate about its potential chirality continues until the present day~\cite{vanwezel11,castellan13, grandhand15, hildebrand18, lin19, rosenkranz19}. Within this context, we establish here that additionally, the electronic ground state in $1T$-TiSe$_2$ has a non-trivial orbital structure.

At the centre of the rich phase diagram and versatility of TMDC materials, but also of their controversies, lie the transition-metal $d$-orbitals. Their spatially compact but not fully localised electronic wave functions allow small changes in the occupation of these orbitals to have a strong effect on the atomic displacement of neighbouring chalcogens, on the distribution of electron charges between atoms, on the coupling of electronic excitations to phonons, and even on the structuring of spins~\cite{tokura00,vanwezel11,Geck15,Silva18,Flicker16}. Unveiling any patterns of orbital occupation therefore offers a novel crucial element for understanding the nature and emergence of order in transition-metal dichalcogenides.

\begin{figure*}[tb]
\centering
\includegraphics[width=\linewidth]{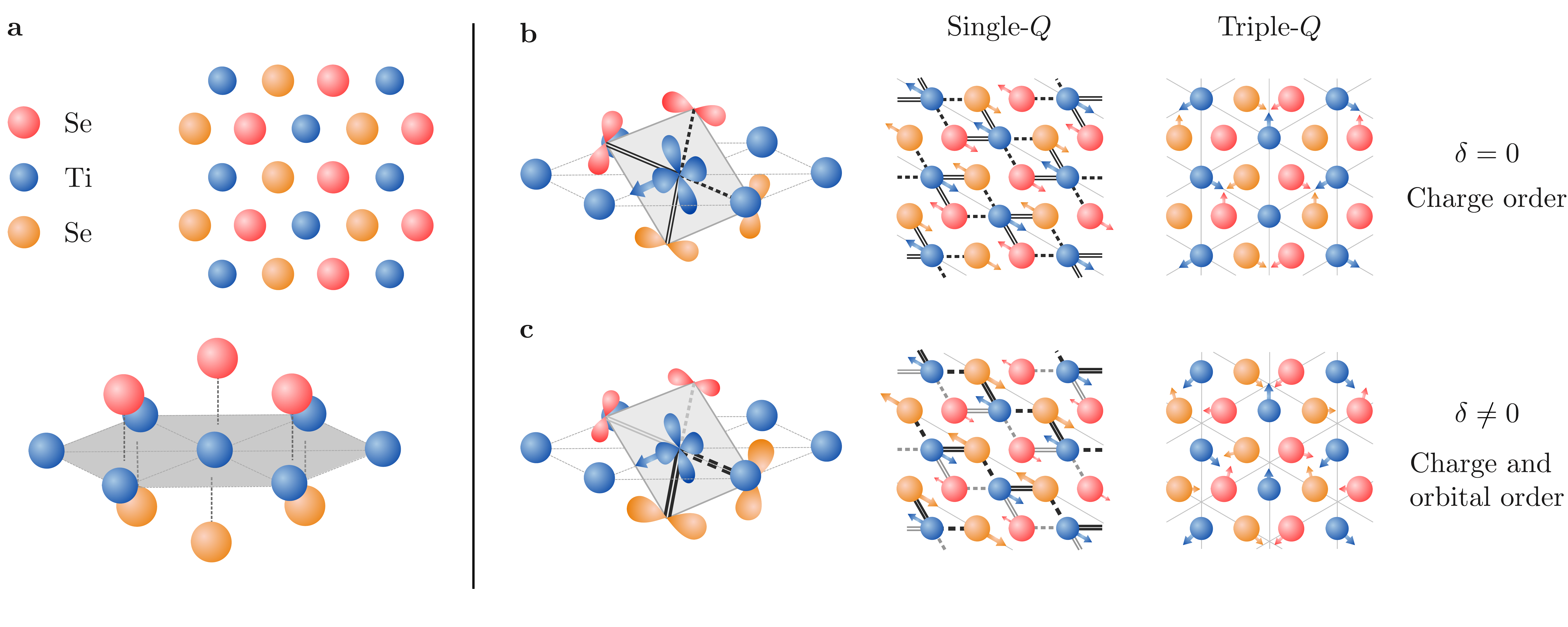}
\caption{{\bf Charge and orbital order in $1T$-TiSe$_2$.} (a) Top-view and three-dimensional sketch of the atomic structure within a single layer of $1T$-TiSe$_2$ in the high-temperature phase (space group 164, $P\bar{3}m1$). (b) The charge ordered phase without relative phase shifts (space group 165, $P\bar{3}c1$). The local charge transfer process for one of the CDW components and all orbitals involved in it are shown in the three-dimensional structure, followed by a top view of the displacements and charge transfers in a single CDW component, and finally the full displacement pattern in the triple-$Q$ structure actually realised in TiSe$_2$. Extrema of the CDW are indicated by faint grey lines and atomic displacements by arrows (exaggerated for clarity). Double lines indicate an electronic bonding state between neighbouring atoms, while dashed lines depict an electronic anti-bonding state. (c) The orbital ordered state resulting from relative phase shifts $\delta$ between CDW components (space group 5, $C2$). The sliding of a single component causes the charge transfer processes to be centered on either the upper or lower Se layer. As a result, the atomic displacements and orbital occupations associated with different CDW components differ in amplitude. This results in a slight alteration to the distortion pattern, as well as an ordered redistribution of electronic charge among the titanium $t_{2g}$ orbitals.
\label{fig1}}
\end{figure*}

That the orbital degree of freedom of electrons in a solid material can be ordered in the same fashion as the electron's charge or spin, was pointed out in the 1970s~\cite{kugel73,kugel82}. Even if electrons in solids form bands and delocalize from the nuclei, in Mott insulators they retain their three fundamental quantum numbers: spin, charge and orbital. This observation sparked a field of research that has gone on to produce a number of important results regarding the cooperative and often concomitant ordering of these degrees of freedom: just as spins can spontaneously organise into regular arrangements and produce myriad types of magnetism, orbital degrees of freedom can also spontaneously order into regular patterns. Such orbital order has been identified using polarised neutron diffraction to measure its effect on the magnetic form factor in for example K$_2$CuF$_4$~\cite{moussa76}, or by measuring how it influences charge distributions in x-ray or electron diffraction in materials like  NdSr$_2$Mn$_2$O$_7$~\cite{takata99}. The first direct observation of orbital order, however, was established for LaMnO$_3$ and La$_{0.5}$Sr$_{1.5}$MnO$_4$ using resonant X-ray scattering (RXS) experiments~\cite{murakami98,staub06}. All of these materials are three-dimensional Mott insulators, characterised by strong Coulomb interactions and correspondingly large variations of orbital occupancy.

Here, we use resonant X-ray scattering measurements at the titanium K-edge to reveal the onset of long-range orbital order among the titanium d-orbitals of the weakly-coupled Van der Waals material $1T$-TiSe$_2$, within its well-known charge ordered phase. The orbitals that order in TiSe$_2$ mediate a strong interaction with spin, charge and lattice degrees of freedom. The orbital order thus not only represents a new ordered phase of matter for VdW materials, but also offers an inroad for engineering complex phase diagrams and devices with beyond-graphene capability.

\section{orbital order in $\text{TiSe}_2$}
The quasi-two-dimensional layers of $1T$-TiSe$_2$ consist of titanium atoms in a triangular arrangement, sandwiched between similar planes of selenium atoms and separated from neighbouring sandwich layers by a large Van der Waals gap, as shown schematically in Fig.~\ref{fig1}a~\cite{whangbo92}. The octahedral coordination of selenium atoms around each titanium splits the titanium d-shell into three $t_{2g}$ and two $e_g$ orbitals. The subsequent hybridization between low-energy $t_{2g}$ orbitals and selenium $p$-orbitals lies at the root of charge and orbital order in TiSe$_2$~\cite{vanwezel11}.

Cooling down from high temperatures, the uniform, metallic state gives way to a well-known charge ordered (CDW) state at $T_{\text{CDW}}\simeq 200$\,K, with atomic displacements (shown in Fig.~\ref{fig1}b) known from neutron diffraction experiments~\cite{disalvo76}. Owing to the lattice symmetry, the charge order in TiSe$_2$ consists of three simultaneous one-dimensional, or single-$Q$, charge density waves, related to one another by $120^{\circ}$ rotations. In each of these, symmetry dictates that electronic charge is transferred within a specific set of titanium $t_{2g}$ and selenium $p$-orbitals, as indicated in Fig.~\ref{fig1}b~\cite{vanwezel11}. The atomic displacements in each of the single-$Q$ CDW components can then be understood as the shortening (extension) of Ti-Se bonds with electronic (anti)bonding states. Notice that as long as the three-fold rotational symmetry of the lattice remains unbroken, all $t_{2g}$ orbitals remain equally occupied.

\begin{figure*}[tb]
    \centering
    \includegraphics[width=\linewidth]{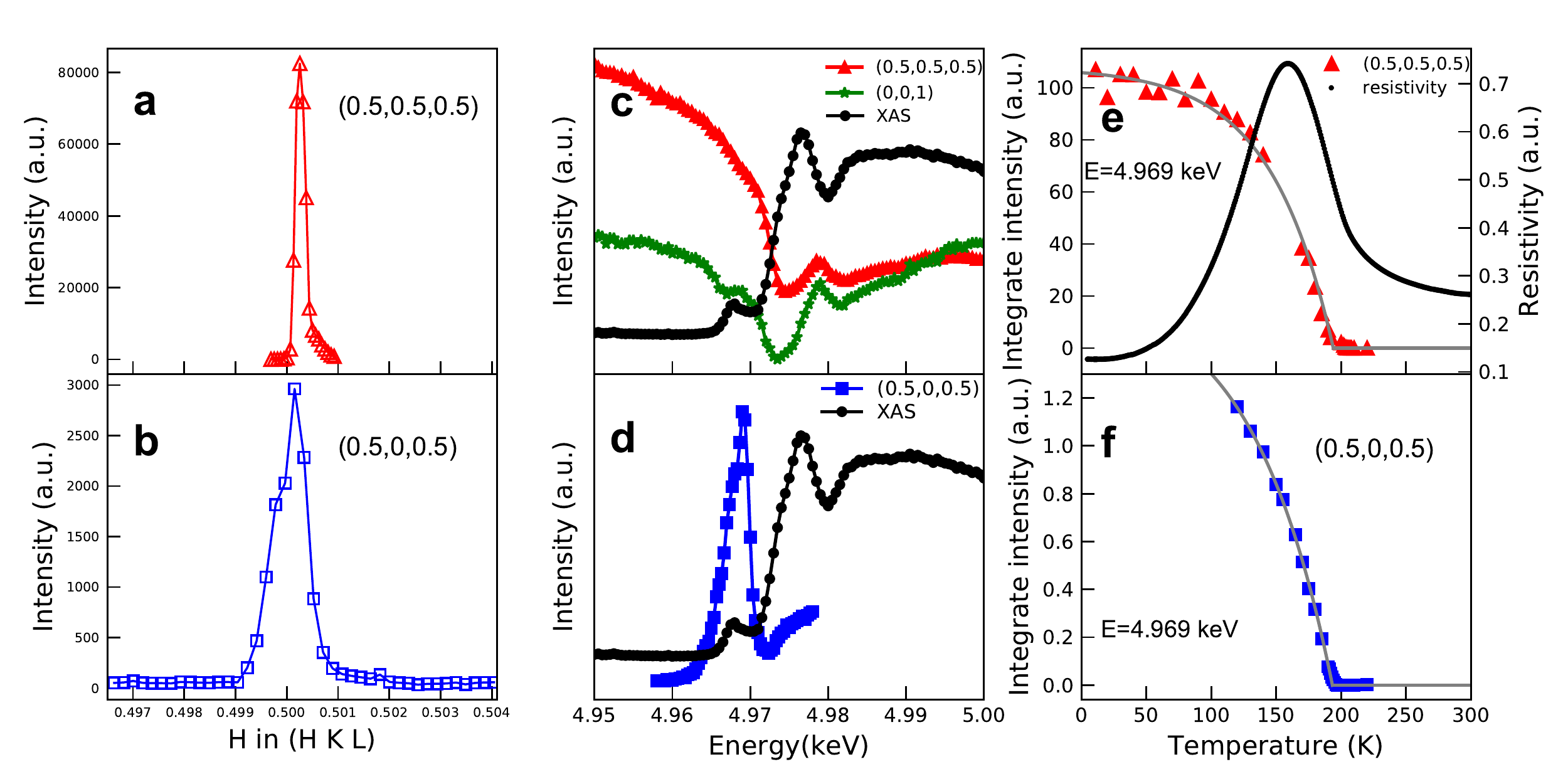}
    \caption{{\bf Charge density wave and orbital order in TiSe$_2$ and their energy- and temperature-dependence measured by RXS at the Ti K-edge.} (a, b) H cuts at 11 K for $(0.5, 0.5, 0.5)$ and $(0.5, 0, 0.5)$, respectively. (c, d) Corresponding energy scans near Ti K-edge. The X-ray Absorption Spectrum (XAS) at the Ti K-edge is shown for comparison, as is the energy scan for the $(0, 0, 1)$ Bragg peak. (e, f) Temperature dependence at 4.969 keV of the peaks at $(0.5, 0.5, 0.5)$ and $(0.5, 0, 0.5)$ respectively. The resistivity measurement is displayed in (e) for comparison. The gray solid lines are fits to the data showing a transition temperature $T_{\text{cdw}}\sim$193\,K. }
    \label{fig2}
\end{figure*}

The three components of the charge ordered state are not independent, and influence one another through local Coulomb interactions. On the basis of theoretical considerations it has been proposed that these yield a second phase transition in TiSe$_2$, in which the three CDW components slide by different amounts in order to minimize the real-space overlap of electronic charge~\cite{vanwezel11,castellan13}. This results in a redistribution of charge between different $t_{2g}$ orbitals, and thus an orbital ordered (OO) state that breaks both the three-fold rotational and inversion symmetries of the high-temperature phases. The atomic displacements associated with the shifting CDW components reflect the broken symmetries, and yield a low-temperature configuration shown in Fig.~\ref{fig1}c. However, the changes in atomic displacement upon entering the OO phase may be expected to be exceedingly small, and evade detection in scanning-tunneling microscopy (STM) and x-ray diffraction experiments~\cite{disalvo76,hildebrand18}.

The actual presence of broken inversion symmetry associated with orbital order in TiSe$_2$ has been the subject of intense debate~\cite{lin19,rosenkranz19,hildebrand18}. On the one hand, there is indirect evidence of a phase transition at temperatures slightly below $T_{\text{CDW}}$~\cite{castellan13}, and STM experiments reporting the formation of domains that appear to have different senses of handedness~\cite{iavarone12}. These are contested however by other groups who observe no evidence of broken inversion symmetry in STM or other probes~\cite{hildebrand18}. The paradox is further fuelled by lack of any direct impact of the breakdown of inversion on scattering experiments, and the practical complication of bulk probes necessarily averaging over many domains of different handedness, thus precluding any direct measurement of broken inversion symmetry. Recently, photogalvanic effect measurements were able to conclusively show inversion symmetry to be broken in the low-temperature phase of samples cooled through their ordering transition in the presence of a strong circularly polarized light field~\cite{xu20}. Because of the active training, however, these results do not rule out the possibility that untrained samples of TiSe$_2$ remain inversion symmetric.

Here, we find that orbital order, rather than the breakdown of inversion symmetry, is actually the main characteristic of the low-temperature phase of TiSe$_2$. The mechanism of CDW components sliding by different amounts under the influence of Coulomb interactions and hence giving rise to spatial modulations in the occupation of orbitals, is generic for multi-component CDW materials, and has been suggested to be at play in the same form in other TMDC compounds, such as $2H$-TaS$_2$, as well as some elemental materials, like Se, Te, and Po~\cite{vanwezel11,vanwezel12,Silva18,Silva18prb}. The RXS experiments presented here, supported by first-principles calculations, show direct experimental evidence for the presence of an orbital ordered phase in TiSe$_2$. Crucially, because the observed signal is insensitive to the handedness of any domains, it is not necessary to train the sample with an applied field, allowing us to detect symmetry breaking that is truly spontaneous.

\section{X-ray scattering}
Orbital order in $3d$ transition-metal compounds can be probed by RXS at the main K-edge corresponding to the $1s\!\rightarrow\!4p$ transition \cite{vedrinskii1998pre,de20091s}. The contribution of Ti $3d$ orbitals to the pre-edge of this transition stems primarily from their hybridization with ligand Se $4p$ orbitals that are in turn hybridized with $4p$ orbitals of neighboring Ti atoms. In addition, lack of inversion symmetry allows for some direct hybridization between Ti $3d$ and $4p$ orbitals at the same site.
Hence, we performed RXS measurements at the Ti K-edge to investigate the ordering
of Ti $3d$ orbitals of $t_{2g}$ symmetry (see Methods section for sample
preparation and experimental set-up). In agreement with RXS measurements at the Se K-edge~\cite{Staub21}, the observed CDW reflections
may be divided into qualitatively different groups, based on their energy
dependence. Unlike the Se-based measurements, however, we find that the different energy profiles of the Ti K-edge reflections allow us to assign them to distinct scattering processes and yield direct evidence for the presence of orbital order. 

The first group of reflections contains conventional CDW peaks,
like the $(0.5,0.5,0.5)$ reflection shown in Fig.~\ref{fig2}a, which are also
visible in non-resonant scattering experiments.
Their energy line shape is similar to that of the primary Bragg peaks, included
for comparison in Fig.~\ref{fig2}c. These signals are dominated by Thomson
scattering from atomic cores, and give a direct measure of the lattice
displacements in the CDW phase. Since all electrons on the Ti atoms participate
in this type of scattering process, it yields a large overall intensity. It
does not have a strong energy dependence, except for a sharp suppression near
the resonant edge, where incoming x-rays can be resonantly absorbed.
Fig.~\ref{fig2}e shows the integrated intensity of the peak in panel a. Fitting
with a universal power-law function (see Methods section for details) yields an
estimated transition temperature of $T_{\text{CDW}} \sim 193$\,K, which is
consistent with resistivity measurements on a sample from the same batch. 

Intriguingly, we identified a second group of RXS signals at points in k-space where reflections are absent in non-resonant experiments, like $(0.5, 0, 0.5)$ shown in Fig.~\ref{fig2}b (another example, $(2.5, 0, 0.5)$, is shown in the Supplemental Material). The structure factor at these momenta is strictly zero as long as the crystal structure of TiSe$_2$ is either $P\bar{3}m1$ or $P\bar{3}c1$, corresponding to the symmetries of the high-temperature phase and the CDW state without orbital order respectively (see Supplemental Material for details of the calculation). The origin for the reflections in the present experiment can be clarified from their energy dependence. Unlike what would be expected from Thomson scattering, these signals have a strong, resonant peak at the pre-edge of the Ti K-edge, as shown in Fig.~\ref{fig2}d. This is typical for reflections originating in transitions between specific orbital states, involving only a few electrons that can be efficiently excited only near resonance~\cite{Yamamoto08}. For this type of reflection, the local density of states dominates the energy dependence of the scattering intensity. 

The presence of a second group of reflections, originating from transitions between specific orbital states, implies that Titanium atoms with the same number of electrons, whose structure factors cancel in Thomson scattering, must have inequivalent orbital structures that allow these reflections under resonant conditions. In other words, the orbital occupations of the Titanium atoms follows a regular pattern with lower symmetry than the high-temperature state and conventional CDW phase. The observation of peaks like $(0.5, 0, 0.5)$ and $(2.5, 0, 0.5)$, with energy dependencies dominated by orbital-specific transitions, thus constitutes direct evidence of orbital order in the low-temperature phase of TiSe$_2$. Based on fits of the thermal evolutions of integrated intensities for these new reflections, shown in Fig.~\ref{fig2}f, we find the onset temperature of the novel orbital ordered state and the conventional CDW phase to be indistinguishable within the experimental resolution.

\begin{figure}[tb]
\centering
\includegraphics[width=\columnwidth]{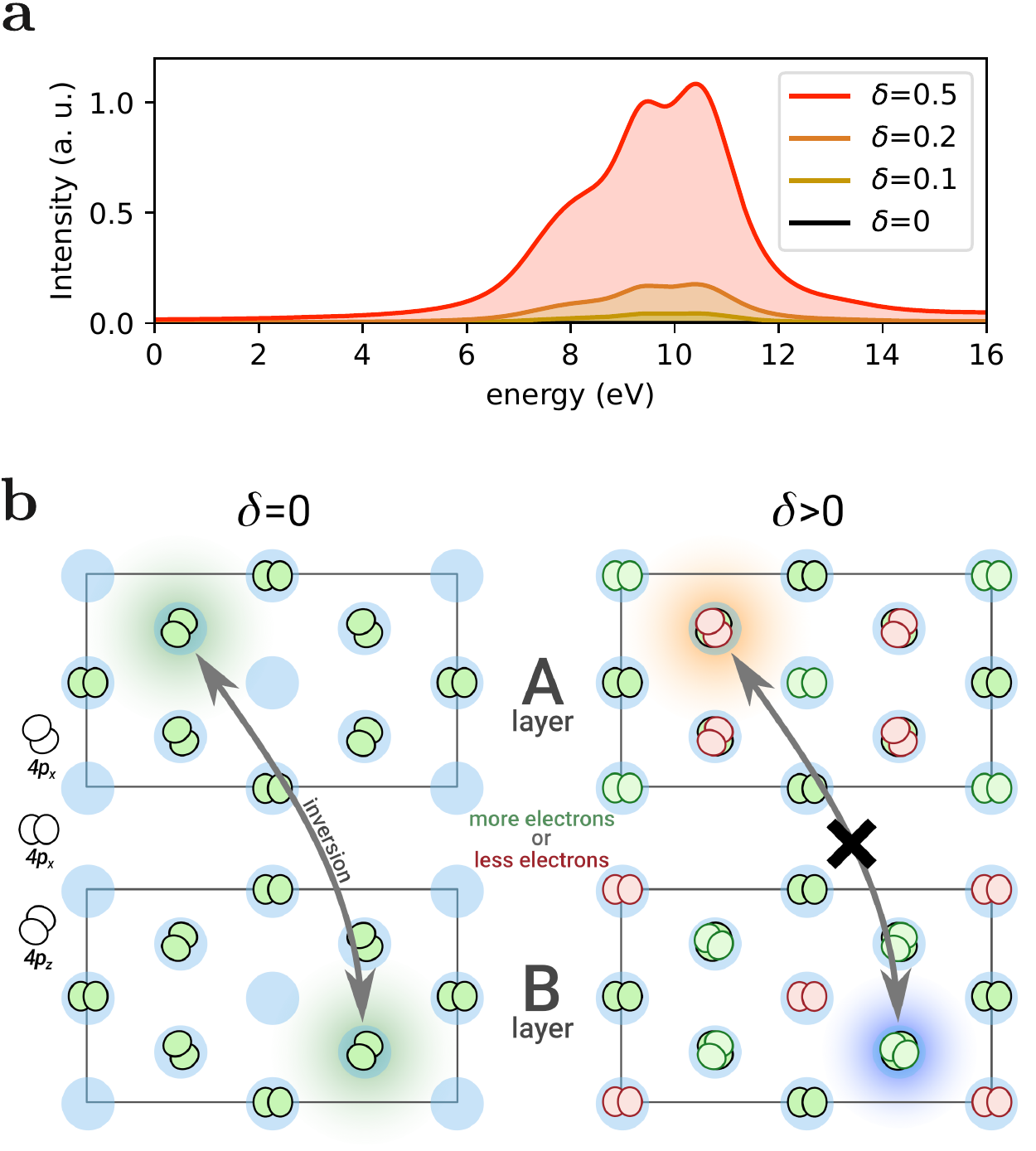}
\caption{
{\bf Energy dependence of the scattering amplitude in the orbital ordered phase.}
(a) Expected energy dependence of the scattering amplitude, with its characteristic resonance structure, based on the square of the orbital-resolved projected densities of states (see Supplemental Material), and broadened to reflect the effect of finite lifetimes. Different curves represent different values of the relative phase shift between CDW components. The energy scale on the horizontal axis is measured with respect to the Fermi level.
(b) Ti $4p$ orbital densities at fixed energy in the atomic structures with zero (left) and non-zero (right) relative phase shift $\delta$ between CDW components. For $\delta=0$, three out of four Ti sites have one predominant $4p$ orbital (green), while orbitals with lower density are degenerate. The structures in neighboring planes are related by inversion. For nonzero relative phase shift, the densities of some previously degenerate orbitals increase (green) or decrease (red). The pattern of orbital polarization shown here for Ti $4p$ orbitals is also present in the densities of other orbitals, and breaks both inversion and three-fold rotational symmetry.
\label{fig3}}
\end{figure}

Although the orbital order is primarily manifested in the occupation of titanium $t_{2g}$ orbitals, it also has an effect on atomic displacements. This is evidenced in yet a third class of reflections that we observe, whose energy dependence indicates a combination of Thomson and orbital-specific scattering (for example, the peak $(0.5, 0.5, 2.5)$ is shown in the Supplemental Material). These reflections are significantly enhanced compared to non-resonant experiments near the Ti K-edge, but also show the largely energy-independent background characteristic of atomic scattering. The enhancement of these reflections can be viewed as a secondary effect of the onset of orbital order, and showcase the strong coupling between orbital, lattice, and charge degrees of freedom.

\section{First-principles results}
A Ginzburg-Landau theory analysis of TiSe$_2$ predicts orbital order to
emerge when different CDW components obtain phase shifts relative to one
another~\cite{vanwezel11}. Adding corresponding phase shifts to the atomic displacements,
gives a model for the atomic configuration of the OO phase (see
Supplemental Material for details). To obtain microscopic insight into
the effects of orbital order, we consider structures with different phase
shifts and calculate projected density of states (PDOS) by means of
density functional theory (DFT) calculations. As resonant peaks are
underlain by uneven orbital occupations~\cite{Elfimov99,Takahashi99,Mahadevan01,benedetti01}, the RXS
structure factor is proportional to a linear combination of the PDOS for
all Ti atoms in the unit cell. In Fig.~\ref{fig3}a, we demonstrate that
a particular linear combination of Ti-$4p$ orbitals reflecting the expected difference in orbital occupation between consecutive layers, yields a nonzero scattering amplitude in structures with nonzero relative phase shifts. 

As shown schematically in Fig.~\ref{fig3}b, the onset of this signal implies differences between the densities of symmetry-related orbitals at fixed energy, and indicates a breakdown of both inversion and three-fold rotational symmetries. 
The presence of orbital order thus allows for resonant reflections that are forbidden in the CDW structure without relative phase shifts. Notice that because TiSe$_2$ is a weakly coupled material, the absolute variations in orbital occupation may be expected to be much lower than those in strongly correlated insulators, such as the archetypical orbitally-ordered material LaMnO$_3$~\cite{murakami98}. Nevertheless, a systematic investigation of which forbidden peaks become allowed and gain significant amplitude in the OO phase (see Supplemental Material), shows that both the momenta at which such reflections emerge, and the energy dependence of the predicted signal, are consistent with the RXS experiments.

\section{Conclusions}
We report experimental evidence for the existence of orbital order in the low-temperature phase of $1T$-TiSe$_2$, consistent with first-principles predictions based on a theoretical model for combined charge and orbital order. 

These observations pave the way for exploring orbital order in other low-dimensional, weakly coupled materials. Based on the general nature of the mechanism stabilising the OO phase, one expects very similar types of orbital order to be common in determining physical properties throughout TMDC with commensurate multi-component charge order, like $2H$-TaS$_2$ or $1T$-VSe$_2$, but also more generally in beyond-graphene Van der Waals materials. 

As all charge transfer processes take place within individual TiSe$_2$ sandwiches, the orbital order is expected to survive even in the monolayer limit. This, together with the strong interplay between orbital occupation and charge, lattice, and spin degrees of freedom, gives great potential to the orbital order as a promising route to tuning and manipulating the physical responses in devices based on Van der Waals materials.

\section{Acknowledgements}
Y.Y.P. is grateful for financial support from the National Natural Science Foundation of China (Grant No. 11974029). P.A. acknowledges support from the Gordon and Betty Moore Foundation, grant no. GMBF-9452. RXS experiments were supported by the U.S. Department of Energy grant no. DE-FG02-06ER46285, with use of the Advanced Photon Source supported by DOE contract no. DE-AC02-06CH11357. H.X.L. acknowledges financial support from the National Natural Science Foundation of China (No. 11922415). J.v.d.B acknowledges the Deutsche Forschungsgemeinschaft (DFG) for support through the W\"urzburg-DresdenCluster of Excellence on Complexity and Topology in Quantum Matter – ct.qmat (EXC 2147, project-id 39085490) and the Collaborative Research Center SFB 1143 (project-id 247310070).

\section{Author contributions}
J.v.d.B., P.A., Y.Y.P. and J.v.W. conceived and designed the experiments and calculations;
Y.Y.P. and X.F.G. performed the RXS experiment at Advanced Photon Source with the help of J.S. and Y.C.. H.X.L and D.Y. grew the TiSe$_2$ samples; S.J and Y.Q.H performed the resistivity measurement; Y.Y.P., Q.X, Q.Z.L and X.F.G analysed the experimental data; J.v.W. modelled atomic displacements and O.J. performed the first-principles calculations; J.v.W. wrote the manuscript with input from all authors.

\section{Methods}
\noindent
{\bf Sample preparation}\\
We synthesised 1T-TiSe$_2$ single crystals in two steps. Firstly, polycrystalline $1T$-TiSe$_2$ powders were synthesized in solid state, by grinding stoichiometric quantities of the reactants of Ti (Alfa Aesar, 99.9\%) and Se (Alfa Aesar, 99.999\%) in an agate mortar. The mixture was then transferred to a quartz tube and sealed under high vacuum. Polycrystalline samples were obtained after heating at 550$^\circ$C for 120 hours. Secondly, single crystals of $1T$-TiSe$_2$ were grown using the chemical vapor transport (CVT) method, with I$_2$ as the transport agent. The as-prepared polycrystalline $1T$-TiSe$_2$ powders were mixed with I$_2$ in a quality ratio of about 20:1. The mixtures sealed in vacuum quartz tubes were then heated for 7 days in a two-zone furnace, where the source and growth zone temperatures were 700$^\circ$C and 600$^\circ$C, respectively. Resistivity measurements using the four-points method were performed on single crystals taken from the same batch of samples as that used for RXS experiments.

\noindent
~\\
{\bf RXS measurements}\\
The RXS experiments were performed at beam line 4-ID-D of the Advanced Photon Source at Argonne National Laboratory. The TiSe$_2$ crystal was mounted on an aluminum sample holder and cooled with a closed-cycle cryostat attached to a four-circle diffractometer. Experiments were done by integrating over all scattered photon energies. The linear polarization was collected with $\sigma$-incident polarization (perpendicular to the scattering plane). The X-ray absorption spectrum (XAS) was obtained by integrating the fluorescence signal. The observed pre-edge feature in these spectra is well-known to be related to the d-electrons~\cite{Yamamoto08}. The temperature dependence of $T_{\text{CDW}}$ can be well-fitted using the empirical function $f(t)=I_0[1-[(t+\alpha)/(1+\alpha))]^{\beta}]$, with $t=T/T_{\text{CDW}}$ the reduced temperature, and $\alpha$, $\beta$, and $T_{\text{CDW}}$ used as fitting parameters~\cite{Joe2014NP}.

\noindent
~\\
{\bf First principles prediction of RXS intensities.}\\
We perform density-functional-theory band-structure calculations using the generalized gradient approximation~\cite{PBE96}, as implemented in the full-potential code FPLO version 18~\cite{FPLO}. Although the structures without relative phase shifts feature a higher symmetry than those with phase shifts, we perform all calculations within the same low-symmetry monoclinic space group $C2$ to ensure that all structures are treated on equal footing. The low symmetry of the unit cell gives rise to numerical noise in the calculated projected DOS, which we
remedy by using a densely sampled $k$-mesh of $41\times{}41\times{}24$ points (20213 points in the irreducible wedge).

\bibliographystyle{naturemag}
\bibliography{biblio}

\beginsupplement
\onecolumngrid

\begin{figure*}[b]
    \centering
    \includegraphics[width=0.9\linewidth]{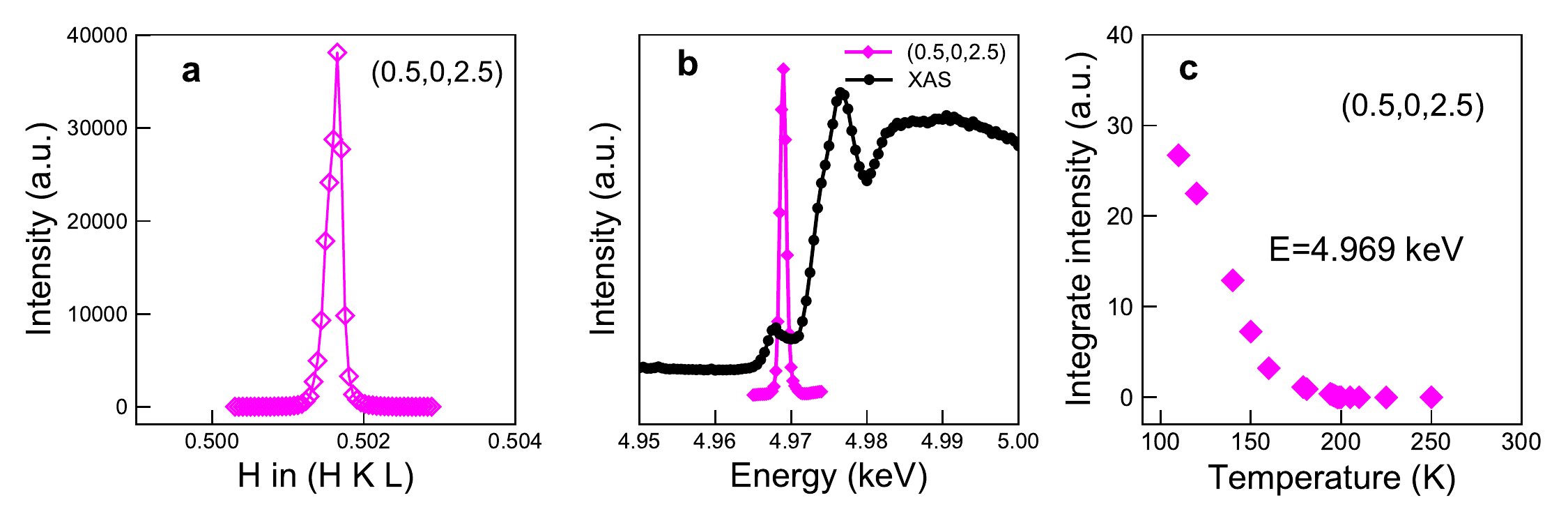}
    \caption{{\bf Orbital order in TiSe$_2$ measured by RXS at the Ti K-edge.} (a) H cut at 11\,K for $(0.5,0,2.5)$. (b) Energy scans near the Ti K-edge at 11\,K. The X-ray absorption spectrum is shown for comparison. (c) Temperature dependence of the integrated intensity for the $(0.5,0,2.5)$ peak taken at 4.969\,keV.}
    \label{SI_fig_1}
    ~\\~\\
    \begin{minipage}{.47\textwidth}
        \centering
        \includegraphics[width=\linewidth]{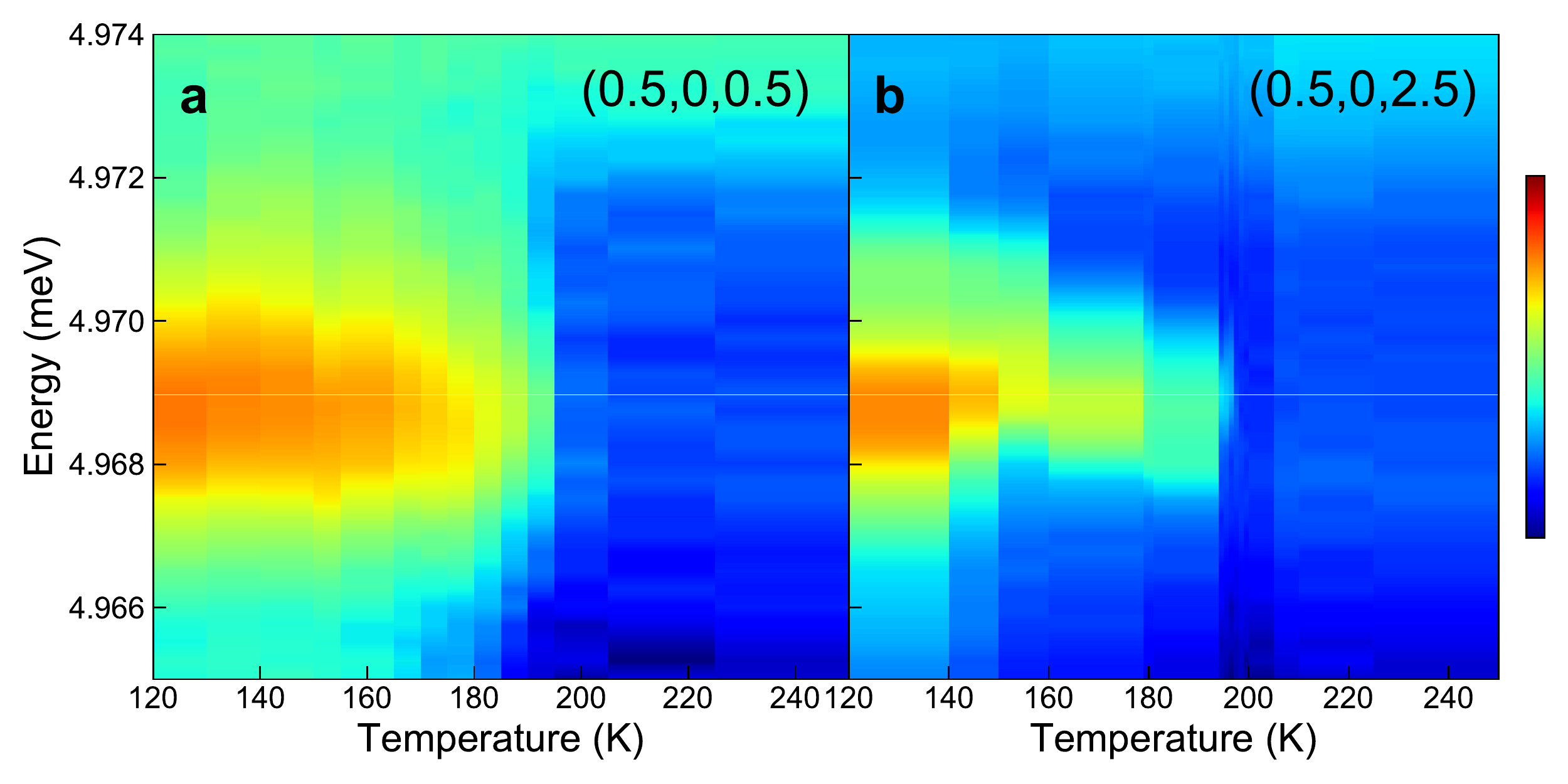}
        \caption{{\bf Temperature dependence of orbital-dominated reflections resonating at the pre-edge of the Ti K-edge.} (a, b) Temperature dependence for energy scans near the Ti K-edge for $(0.5,0,0.5)$ and $(0.5,0,2.5)$, respectively.}
        \label{SI_fig_2}
    \end{minipage}%
    \hspace{0.05\textwidth}
    \begin{minipage}{.47\textwidth}
        \centering
        \includegraphics[width=\linewidth]{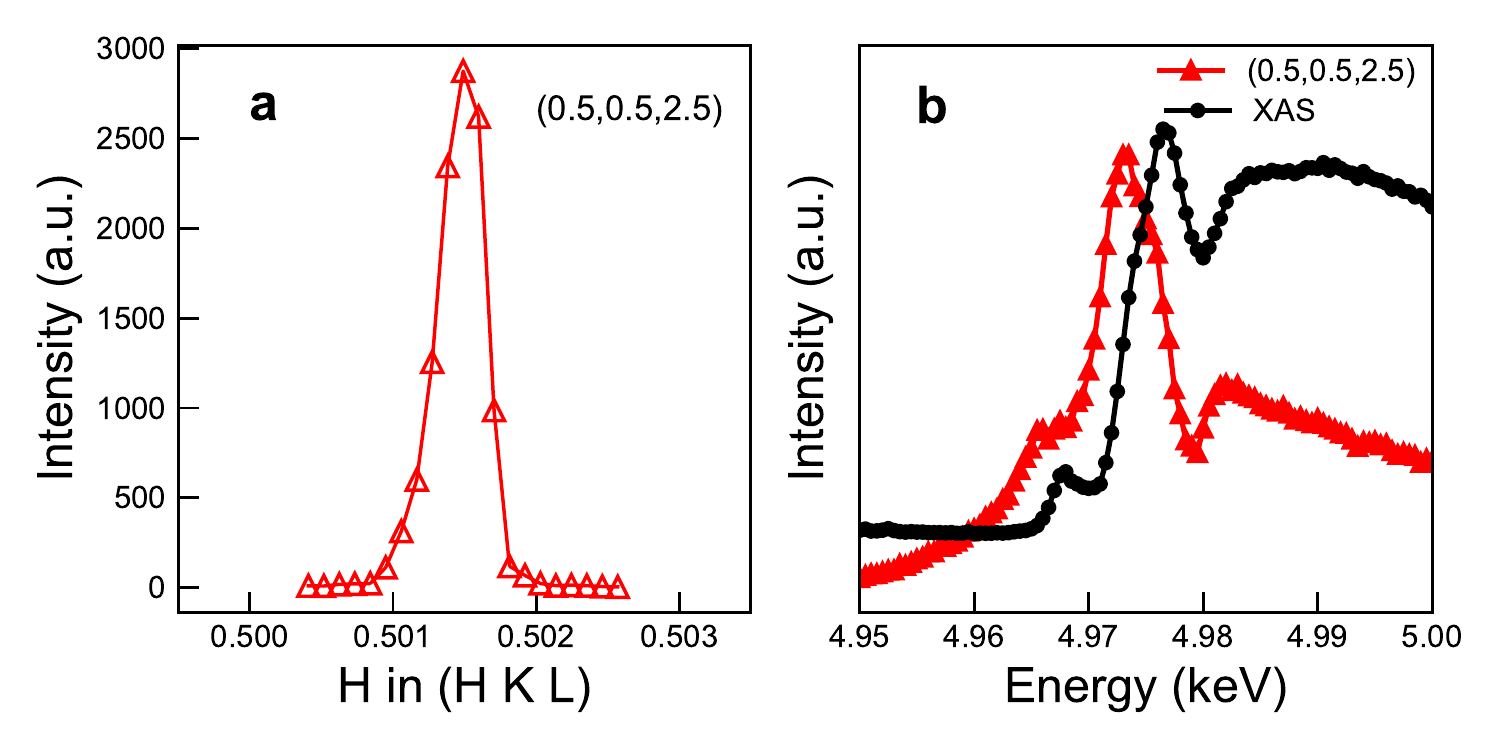}
        \caption{{\bf Charge and orbital order in TiSe$_2$ measured by RXS at the Ti K-edge.} (a) H cut at 11\,K for (0.5,0.5,2.5). (b) Energy scans near the Ti K-edge. The X-ray absorption spectrum near the Ti K-edge is shown for comparison.}
        \label{SI_fig_3}
    \end{minipage}
\end{figure*}

\section{Supplemental Material}
\noindent
The supplementary information below includes details of RXS reflections and dichroism experiments not discussed in the main text, as well as the calculated structure factors for reflections that are forbidden in any configuration with simultaneous inversion and three-fold rotational symmetries.

\noindent
~\\
{\bf X-ray scattering}\\
Besides the $(0.5,0,0.5)$ reflection discussed in the main text, the orbital order in TiSe$_2$ can also be seen at other reflections forbidden in inversion and three-fold rotationally symmetric phases, such as the $(0.5,0,2.5)$ reflection shown in Fig.~\ref{SI_fig_1}a. Similar to the $(0.5,0,0.5)$ reflection, Fig.~\ref{SI_fig_1}b shows this peak to resonate at the pre-edge of the Ti K-edge, indicating its origin in variations of the occupation of Ti $d$-orbitals. Its onset temperature is $\sim$193\,K (Fig.~\ref{SI_fig_1}c), similar to that of the $(0.5,0,0.5)$ signal. This is confirmed by energy scans near the Ti K-edge as a function of temperature, shown in Fig.~\ref{SI_fig_2}, indicating that the resonant behaviour for both reflections disappears above 193\,K.

Besides the two classes of peaks discussed in the main text, one dominated by orbital order and the other by lattice displacements, we also observe a third group of reflections. An example of this is the peak at $(0.5, 0.5, 2.5)$, shown in Fig.~\ref{SI_fig_3}. It resonates near the Ti K-edge rather than the pre-edge, and has an energy profile intermediate between the orbital and lattice dominated reflections. This is indicative of this peak originating from a combination of lattice and orbital scattering.

\begin{figure*}[tb]
    \centering
    \includegraphics[width=1\linewidth]{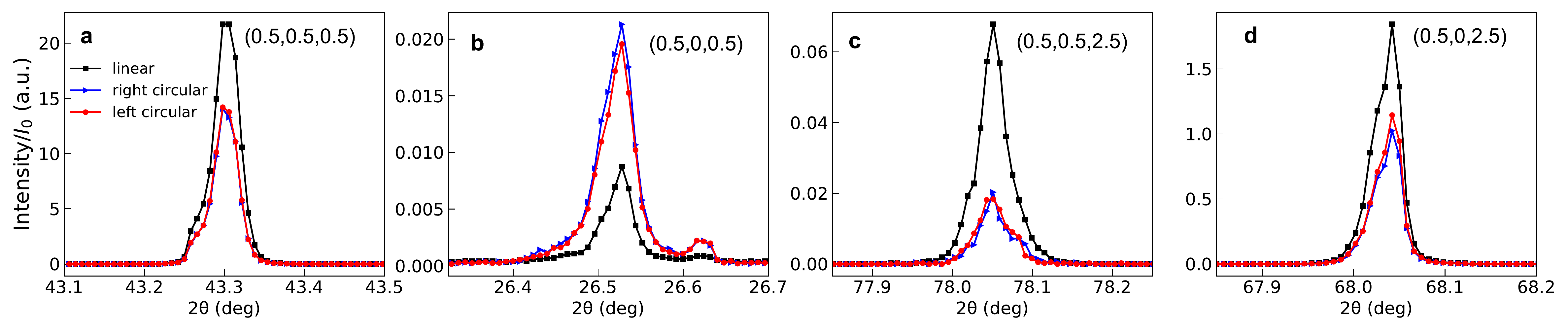}
    \caption{{\bf Polarization dependence of four reflections.} (a, b, c, d) $\theta$-$2\theta$ scans taken at 4.969 keV with linear, right circular, and left circular polarizations for $(0.5,0.5,0.5)$, $(0.5,0,0.5)$, $(0.5,0.5,2.5)$ and $(0.5,0,2.5)$, respectively.}
    \label{SI_fig_4}
\end{figure*}

Finally, we explored the dependence of the four reflections $(0.5,0.5,0.5)$, $(0.5,0,0.5)$, $(0.5,0.5,2.5)$, and $(0.5,0,2.5)$ on the incident polarization, as shown in Fig.~\ref{SI_fig_4}. We considered both linear and circular polarization, and our measurement were done without an analyser resolving the polarization of the scattered signal. The data were normalized by the incident X-ray flux ($I_0$) for comparison. We did not find clear evidence for dichroism in these experiments. The side peaks or shoulders are due to small crystal mosaicity.

\noindent
\\
{\bf Atomic structure}\\
The atomic positions used in the first-principles calculations discussed in the
main text are listed in Table~\ref{tab:coordinates}. Here, $\pm z_{\text{Se}}$
is the fractional height of the Se atoms above and below the Ti plane. The CDW
amplitude $A$ is non-zero in the charge ordered phases, while $\gamma$ is the
ratio of Se to Ti electron-phonon coupling strengths. The parameter $\delta$
finally, represents a relative phase shift of the three CDW components, and is
nonzero in the orbital-ordered phase.

\begin{table}[b!]
\begin{tabular}{rrrr}
    \hline\hline
    atom & $x/a$ & $y/b$ & $z/c$ \\ \hline
    Ti(1) & 0 & 0 & 0 \\
    Ti(2) & $\frac{\sqrt{3}}{2}( 1 + 2A)$ & $\frac12(-1 - 2A)$ & 0 \\
    Ti(3) & $\frac{\sqrt{3}}{2}(-1 - 2A)$ & $\frac12(-1 - 2A)$ & 0 \\
    Ti(4) & 0 & $1 + 2A$ & 0 \\

    Se(1)& $\frac{1}{\sqrt{3}} \left( 1 - \frac32\delta\gamma{}A\right)$ & $2\gamma{}A + \frac12\delta\gamma{}A$ & $z_{\text{Se}}$ \\
    Se(2)& $\frac{1}{2\sqrt{3}}\left(-1 - 6\gamma{}A + 3\delta\gamma{}A\right)$ & $ \frac12(1 - 2\gamma{}A - d\gamma{}A)$ & $z_{\text{Se}}$ \\
    Se(3)& $\frac{1}{2\sqrt{3}}\left(-1 + 6\gamma{}A - 3\delta\gamma{}A\right)$ & $\frac12\left(-1 - 2\gamma{}A - 3\delta\gamma{}A\right)$ & $z_{\text{Se}}$ \\
    Se(4)& $\frac{2}{\sqrt{3}} \left(-1 + \frac34\delta\gamma{}A\right)$ & $\frac32\delta\gamma{}A$ & $z_{\text{Se}}$ \\
    Se(5)& $\frac{1}{\sqrt{3}} \left(-1 + \frac32\delta\gamma{}A\right)$ & $2\gamma{}A - \frac12\delta\gamma{}A$ & $-z_{\text{Se}}$ \\
    Se(6)& $\frac{1}{2\sqrt{3}}\left( 1 - 6\gamma{}A - 3\delta\gamma{}A\right)$ & $ \frac12\left(-1 - 2\gamma{}A + d\gamma{}A\right)$ & $-z_{\text{Se}}$ \\
    Se(7)& $\frac{1}{2\sqrt{3}}\left( 1 + 6\gamma{}A + 3\delta\gamma{}A\right)$ & $ \frac12\left( 1 - 2\gamma{}A +   3\delta\gamma{}A\right)$ & $-z_{\text{Se}}$ \\
    Se(8)& $\frac{2}{\sqrt{3}} \left(1 - \frac34\delta\gamma{}A\right)$ & $-\frac32\delta\gamma{}A$ & $-z_{\text{Se}}$ \\ \hline\hline
\end{tabular}
\caption{\label{tab:coordinates} Fractional atomic coordinates of a single
TiSe$_2$ layer as a function of the structural parameters $A$, $\gamma$,
$\delta$, and $z_{\text{Se}}$. Adjacent layers of TiSe$_2$ have the same
positions, but with $A$ switching sign every layer.} \end{table}

The high-temperature pristine phase with $A=\delta=0$ has space group
$P\bar{3}m1$ (number 164). The symmetry is lowered to $P\bar{3}c1$ (number 165)
when $A$ is nonzero but $\delta$ remains zero. Notice that any distortions of
this state that respect both inversion and three-fold rotational symmetry yield
zero structure factors for reflections like $(0.5,0,0.5)$ and $(0.5,0,2.5)$. If
$\delta$ becomes nonzero, both three-fold rotations and inversion symmetry are
broken, and the space group reduces to $C2$ (number 5).

\begin{figure*}[t]
    \includegraphics[width=0.8\textwidth]{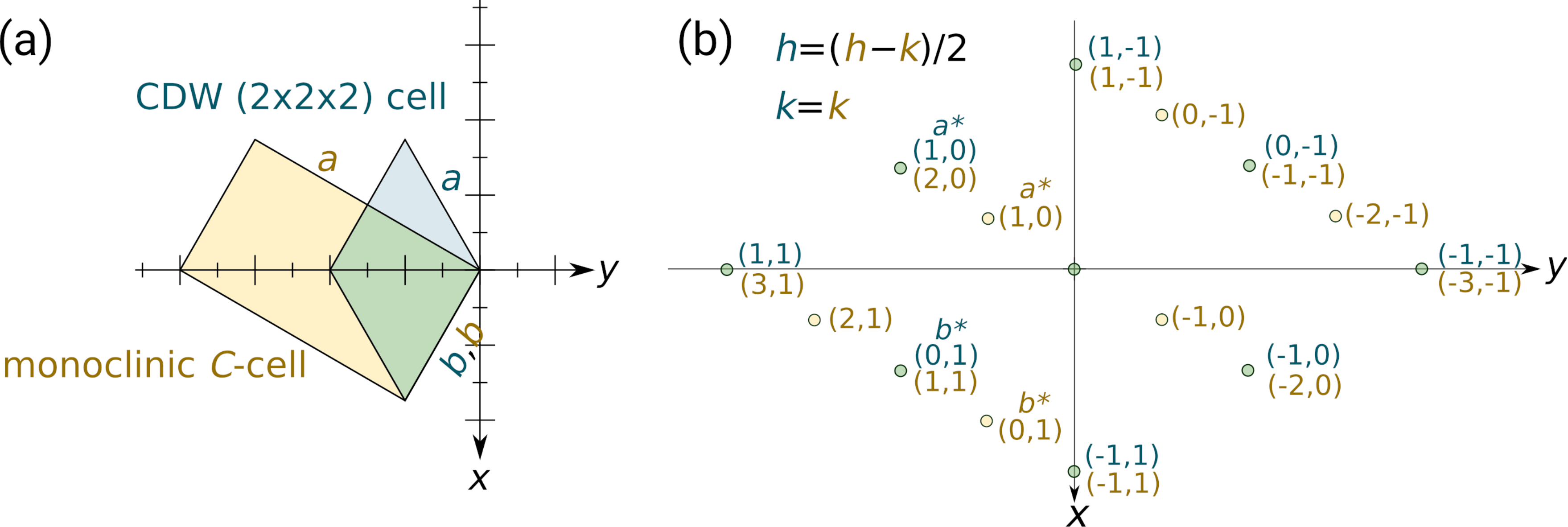}
    \caption{{\bf Comparison of unit cell geometries} (a) Real-space sketch of the unit cells associated with the charge (trigonal) and orbital ordered (monoclinic) phases. (b) Comparison of the corresponding momentum-space indices.}
    \label{SI_cells}
\end{figure*}

In Table~\ref{tab:Shkl}, we list some of the peaks for which the combination of
three-fold rotational and inversion symmetries dictates a zero structure factor
in the high-temperature and charge ordered phases. For completeness, they are
given in notation based on the trigonal unit cell for the high-temperature
phase, as well as with respect to unit cells with the symmetries of the charge
and orbital ordered phases. The sketch in Fig.~\ref{SI_cells} shows a
comparison between the different unit cells and their momentum-space indices.

\begin{table*}[b]
\begin{tabular}{rp{2em}rrrp{2em}rrrrp{2em}rrrr} \hline \hline
& & \multicolumn{3}{l}{trigonal unit cell} & & \multicolumn{4}{l}{enlarged trigonal unit cell} & &\multicolumn{4}{l}{monoclinic unit cell} \\
& & \multicolumn{3}{l}{$P\bar{3}m1$ (164)} & & \multicolumn{4}{l}{$P\bar{3}c1$ (165)} & &\multicolumn{4}{l}{$C2$ (5)} \\
& & \multicolumn{4}{l}{\phantom{enlarged trigonal unit cell}} & \multicolumn{4}{l}{} & &\multicolumn{4}{l}{} \\
& & \multicolumn{3}{l}{$a$ = 3.535\,\AA} & & \multicolumn{4}{l}{$a$ = 7.07\,\AA} & &\multicolumn{4}{l}{$a$ = 12.2456\,\AA} \\
& & \multicolumn{3}{l}{} & & \multicolumn{4}{l}{} & &\multicolumn{4}{l}{$b$ = 7.07\,\AA} \\
& & \multicolumn{3}{l}{$c$ = 6.01\,\AA} & & \multicolumn{4}{l}{$c$ = 12.02\,\AA} & &\multicolumn{4}{l}{$c$ = 12.02\,\AA} \\
& & \multicolumn{3}{l}{} & & \multicolumn{4}{l}{} & &\multicolumn{4}{l}{$\beta$ = 90$^{\circ}$} \\ \hline
{$d$, \AA} & & $h$ & $k$ & $l$ & & $h$ & $k$ & $l$ & & & $h$ & $k$ & $l$ \\ \hline
 12.02000 & &$ 0  $ &$ 0  $ &$\pm\nicefrac12$  & &$ 0  $ &$ 0$ &\,$\pm1$&   & &$ 0$ &$ 0$ &\,$\pm1$& \\
  5.45577 & &$ \nicefrac12$ &$ 0  $ &$\pm\nicefrac12$  & &$ 1  $ &$ 0$ &$\pm1$&   & &$ 2$ &$ 0$ &$\pm1$& \\
  5.45577 & &$ 0  $ &$ \nicefrac12$ &$\pm\nicefrac12$  & &$ 0  $ &$ 1$ &$\pm1$&   & &$ 1$ &$ 1$ &$\pm1$& \\
  5.45577 & &$ \nicefrac12$ &$-\nicefrac12$ &$\pm\nicefrac12$  & &$ 1  $ &$-1$ &$\pm1$&   & &$ 1$ &$-1$ &$\pm1$& \\
  5.45577 & &$-\nicefrac12$ &$ \nicefrac12$ &$\pm\nicefrac12$  & &$-1  $ &$ 1$ &$\pm1$&   & &$-1$ &$ 1$ &$\pm1$& \\
  5.45577 & &$ 0  $ &$-\nicefrac12$ &$\pm\nicefrac12$  & &$ 0  $ &$-1$ &$\pm1$&   & &$-1$ &$-1$ &$\pm1$& \\
  5.45577 & &$-\nicefrac12$ &$ 0  $ &$\pm\nicefrac12$  & &$-1  $ &$ 0$ &$\pm1$&   & &$-2$ &$ 0$ &$\pm1$& \\
  4.00667 & &$ 0  $ &$ 0  $ &$\pm\nicefrac32$  & &$ 0  $ &$ 0$ &$\pm3$&   & &$ 0$ &$ 0$ &$\pm3$& \\
  3.35263 & &$ \nicefrac12$ &$ 0  $ &$\pm\nicefrac32$  & &$ 1  $ &$ 0$ &$\pm3$&   & &$ 2$ &$ 0$ &$\pm3$& \\
  3.35263 & &$ 0  $ &$ \nicefrac12$ &$\pm\nicefrac32$  & &$ 0  $ &$ 1$ &$\pm3$&   & &$ 1$ &$ 1$ &$\pm3$& \\
  3.35263 & &$ \nicefrac12$ &$-\nicefrac12$ &$\pm\nicefrac32$  & &$ 1  $ &$-1$ &$\pm3$&   & &$ 1$ &$-1$ &$\pm3$& \\
  3.35263 & &$-\nicefrac12$ &$ \nicefrac12$ &$\pm\nicefrac32$  & &$-1  $ &$ 1$ &$\pm3$&   & &$-1$ &$ 1$ &$\pm3$& \\
  3.35263 & &$ 0  $ &$-\nicefrac12$ &$\pm\nicefrac32$  & &$ 0  $ &$-1$ &$\pm3$&   & &$-1$ &$-1$ &$\pm3$& \\
  3.35263 & &$-\nicefrac12$ &$ 0  $ &$\pm\nicefrac32$  & &$-1  $ &$ 0$ &$\pm3$&   & &$-2$ &$ 0$ &$\pm3$& \\
  2.96669 & &$ 1  $ &$ 0  $ &$\pm\nicefrac12$  & &$ 2  $ &$ 0$ &$\pm1$&   & &$ 4$ &$ 0$ &$\pm1$& \\
  2.96669 & &$-1  $ &$ 0  $ &$\pm\nicefrac12$  & &$-2  $ &$ 0$ &$\pm1$&   & &$-4$ &$ 0$ &$\pm1$& \\
  2.96669 & &$ 0  $ &$ 1  $ &$\pm\nicefrac12$  & &$ 0  $ &$ 2$ &$\pm1$&   & &$ 2$ &$ 2$ &$\pm1$& \\
  2.96669 & &$ 1  $ &$-1  $ &$\pm\nicefrac12$  & &$ 2  $ &$-2$ &$\pm1$&   & &$ 2$ &$-2$ &$\pm1$& \\
  2.96669 & &$-1  $ &$ 1  $ &$\pm\nicefrac12$  & &$-2  $ &$ 2$ &$\pm1$&   & &$-2$ &$ 2$ &$\pm1$& \\
  2.96669 & &$ 0  $ &$-1  $ &$\pm\nicefrac12$  & &$ 0  $ &$-2$ &$\pm1$&   & &$-2$ &$-2$ &$\pm1$& \\
  2.43259 & &$ 1  $ &$ 0  $ &$\pm\nicefrac32$  & &$ 2  $ &$ 0$ &$\pm3$&   & &$ 4$ &$ 0$ &$\pm3$& \\
  2.43259 & &$ 0  $ &$ 1  $ &$\pm\nicefrac32$  & &$ 0  $ &$ 2$ &$\pm3$&   & &$ 2$ &$ 2$ &$\pm3$& \\
  2.43259 & &$ 1  $ &$-1  $ &$\pm\nicefrac32$  & &$ 2  $ &$-2$ &$\pm3$&   & &$ 2$ &$-2$ &$\pm3$& \\
  2.43259 & &$-1  $ &$ 1  $ &$\pm\nicefrac32$  & &$-2  $ &$ 2$ &$\pm3$&   & &$-2$ &$ 2$ &$\pm3$& \\
  2.43259 & &$ 0  $ &$-1  $ &$\pm\nicefrac32$  & &$ 0  $ &$-2$ &$\pm3$&   & &$-2$ &$-2$ &$\pm3$& \\
  2.43259 & &$-1  $ &$ 0  $ &$\pm\nicefrac32$  & &$-2  $ &$ 0$ &$\pm3$&   & &$-4$ &$ 0$ &$\pm3$& \\ \hline \hline
\end{tabular}
\caption{\label{tab:Shkl} Reflections whose structure factor is strictly zero in the high-temperature and charge ordered phases, but which become observable in the orbital ordered state. The three sets of indices are relevant to the unit cells of the high-temperature, charge-ordered, and orbital ordered phases.}
\end{table*}

\clearpage
\noindent
{\bf Structure factors}\\
With the atomic positions in Table~\ref{tab:coordinates}, we can calculate the
static structure factor, $S(\vec{k})$, which determines the scattering
intensity in X-ray experiments. The contribution of a single atom at position
$\vec{r}$ to the structure factor is: \begin{align}
    S_{\vec{r}}(\vec{k}) = f(\vec{k}) e^{i\,\vec{k}\cdot\vec{r}}.
    \label{Sdef}
\end{align}
Here, $f(\vec{k})$ is the atomic form factor for the atom at position
$\vec{r}$, which varies for different species of atoms. For spherically
symmetric atoms, the form factor depends only on the magnitude of the momentum.
The total structure factor that determines the experimental response is the sum
of atomic structure factors for all atoms in the unit cell. In resonant
scattering experiments, the selective sensitivity to only a single intra-atomic
transition implies that resonantly enhanced signals are dominated by only the
contributions to the structure factor coming from a single species of atom. In
the presence of orbital order, variations of orbital occupation moreover
influence the number of electrons that can be resonantly excited between
specific orbitals, effectively causing the atomic form factors for atoms with
different orbital content to vary.

For a non-resonant scattering experiment, we can approximate all atomic form
factors to be determined solely by the atomic species they refer to. Inserting
the atomic positions (Table~\ref{tab:coordinates}) of either all Ti atoms or
all Se atoms into Eq.~\eqref{Sdef}, the contribution of each atomic species to
the structure factor can be determined. As long as the parameter $\delta$ is
zero, we find that for any value of $A$, the total structure factor for peaks
of the type $(h,0,l)$ with $h$ and $l$ both half-integer is zero. The same
holds for peaks of the type $(0,k,l)$ and $(h,-h,l)$ with $h$, $k$, and $l$ all
half-integer. In other words, reflections with these indices are forbidden in
any structure with the space group $P\bar{3}c1$ associated with the charge
ordered phase. 

For atomic displacements that alter the space group, including those
corresponding with non-zero $\delta$, forbidden reflections may become allowed.
The fact that this class of reflections is not observed in any non-resonant
experiments is thus consistent with an atomic configuration in (or very
close to) that of the charge ordered phase.

\begin{figure*}[t]
    \includegraphics[width=0.5\textwidth]{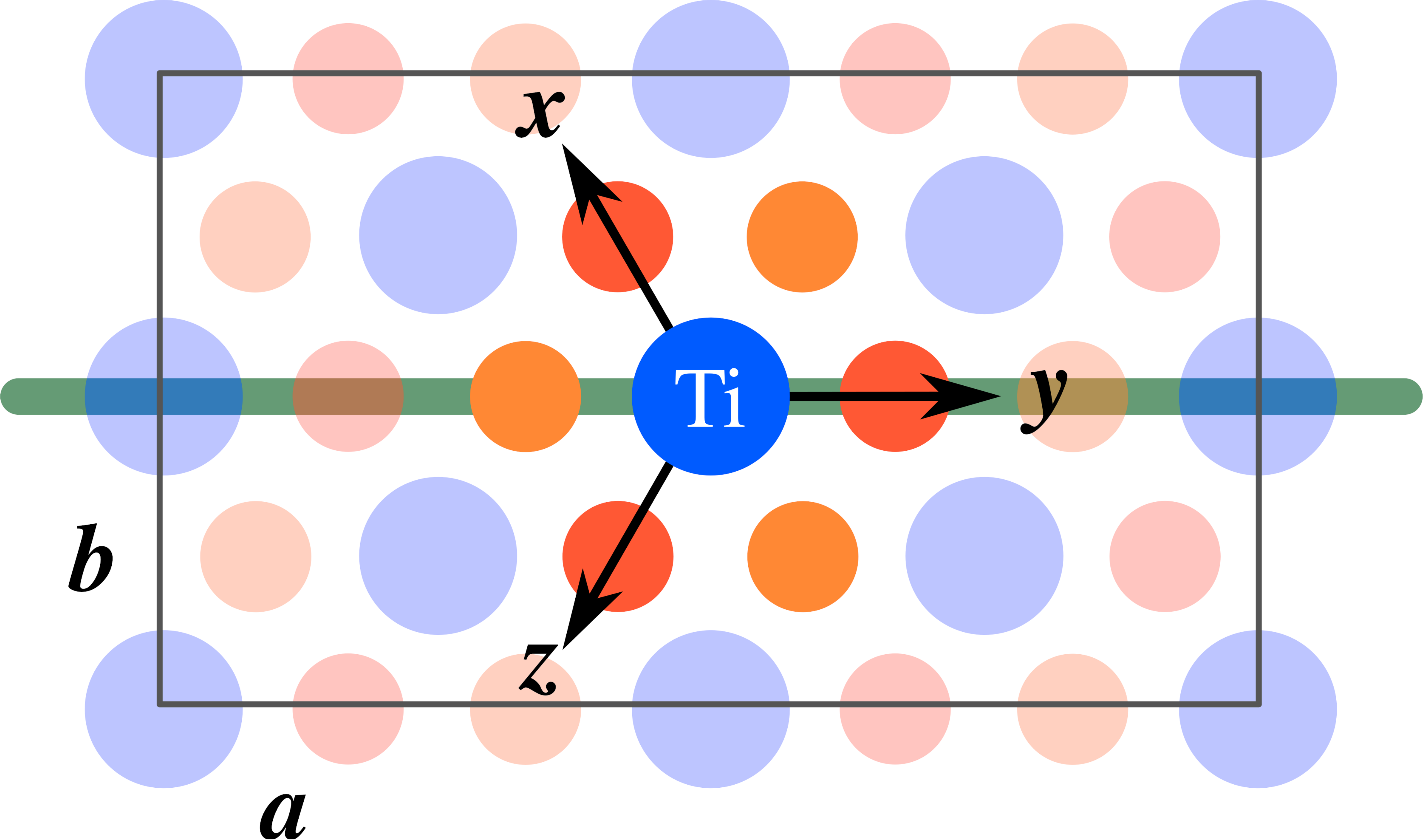}
    \caption{{\bf Choice of the local coordinate frame} The grey rectangle
indicates the boundary of the monoclinic unit cell (space group C2, number 5).  The green line indicates the plane containing the incident and scattered beams, which coincides with the projection of the local $y$ axis onto the $ab$ plane.}
    \label{SI_axes}
\end{figure*}

Turning to resonant scattering experiments next, we can model the effect of
variations in orbital occupation by assigning different atomic form factors to
atoms whose local density of states is predicted to be different in our
first-principles calculation. This yields nonzero scattering intensities for
all of the peaks listed in Table~\ref{tab:Shkl}, which are forbidden
reflections in the absence of orbital order.

\noindent
\\
{\bf Projected Density of States}\\
For calculating the projected DOS of Ti $3d$ and $4p$ states, we use a local coordinate system with axes schematically illustrated in Fig.~\ref{SI_axes}. Their directions in terms of orthogonal coordinates aligned with the crystallographic $a$ and $c$ axes in the high-temperature phase is given by $x=(-\sqrt{1/3}; 1; -\sqrt{2/3})$ and $z=(-\sqrt{1/3}; -1; -\sqrt{2/3})$. Note that we will also use $3d$ and $4p$ orbitals defined with respect to this local
coordinate frame, rather than the crystallographic axes. Focusing on the azimuthal angle such that the $a$ axis of the monoclinic unit cell lies in the
plane formed by the incident and the scattered beam, the relevant projected
Ti $4p$ and $3d$ orbital densities on atom $j$, $D_j(4p)$ and $D_j(3d)$, are given by:

\begin{align}
\label{eqn:pdos_individual}
D_j(3d) &= D_j(3d_{xz}) + e^{i\frac23\pi{}}D_j(3d_{3z^2-r^2}) + e^{i\frac43\pi{}}D_j(3d_{yz}) \notag \\
D_j(4p) &= D_j(4p_y) + e^{i\frac23\pi{}}D_j(4p_z) + e^{i\frac43\pi{}}D_j(4p_x).
\end{align}

The RXS structure factor for reflections that originate from orbital transitions rather than Thomson scattering in the low temperature phase and which are forbidden in the high temperature phase, is proportional to the difference between the orbital occupations on atoms whose locations are related by the inversion symmetry of the high temperature structure~\cite{Elfimov99,Takahashi99,Mahadevan01,benedetti01}. The following linear combination $D$ of projected DOS $D_j$ combines all such pairs:

\begin{equation}
\label{eqn:pdos_combination}
D = D_{\text{Ti(1)A}} + D_{\text{Ti(4)A}} + e^{i\frac23\pi{}}D_{\text{Ti(2)A}} + e^{i\frac43\pi{}}D_{\text{Ti(3)A}} \\
    -D_{\text{Ti(1)B}} -D_{\text{Ti(4)B}} - e^{i\frac23\pi{}}D_{\text{Ti(2)B}} - e^{i\frac43\pi{}}D_{\text{Ti(3)B}}.
\end{equation}
Here, A and B refer to the two layers within the unit cell, with for instance
$D_{\text{Ti(4)A}}$ denoting the projected DOS for atom Ti(4) in layer
A. Using Eqns.~\ref{eqn:pdos_individual} and \ref{eqn:pdos_combination}, we can 
qualitatively estimate the RXS structure factor directly from the atom and orbital
resolved DOS obtained in density-functional-theory
band-structure calculations.

\begin{figure*}[t]
    \includegraphics[width=\textwidth]{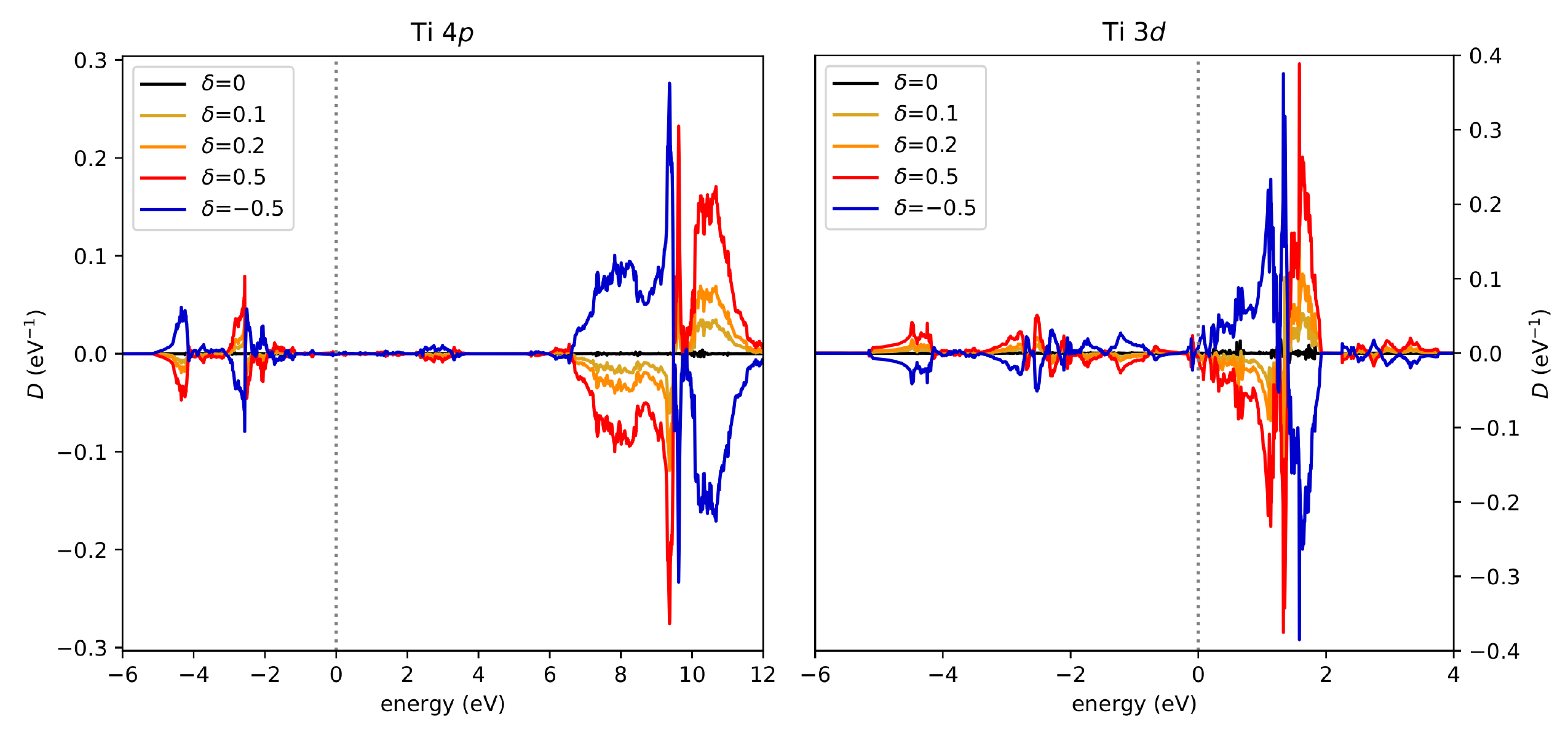}
    \caption{\textbf{Projected Density of States} The linear combinations of Ti $4p$ (left) and $3d$ (right) projected DOS, as given by Eq.~\ref{eqn:pdos_combination}. The Fermi level is at zero energy. Small non-zero values of projected DOS for $\delta=0$ are due to numerical noise.}
    \label{SI_pdos}
\end{figure*}

The resulting $D(3d)$ and $D(4p)$ are shown in Fig.~\ref{SI_pdos}. The small nonzero contributions for the case $\delta=0$ have been verified to be numerical artifacts by confirming that their absolute values systematically decrease for denser $k$-meshes. Comparisons of the results for $\delta=0.1$, $\delta=0.2$ and $\delta=0.5$ shows that the projected DOS increases systematically with $\delta$.  As the sign of the parameter $\delta$ is not fixed, we performed calculations also for the $\delta=-0.5$ case. The resulting plots are reflections in the $D=0$ axis of the corresponding $\delta=0.5$ curves.

\begin{figure*}[t]
  \includegraphics[width=.49\textwidth]{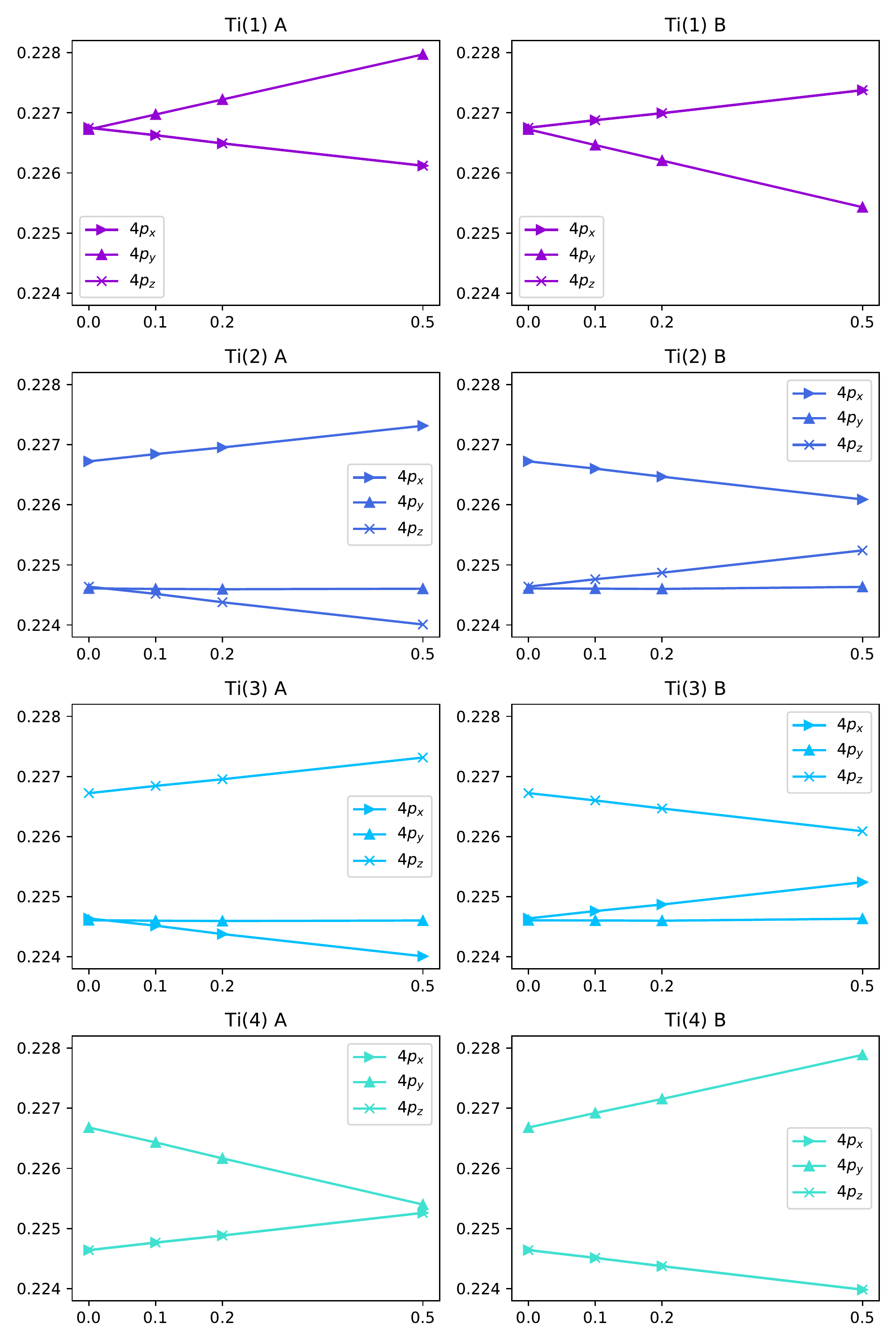}
  \includegraphics[width=.49\textwidth]{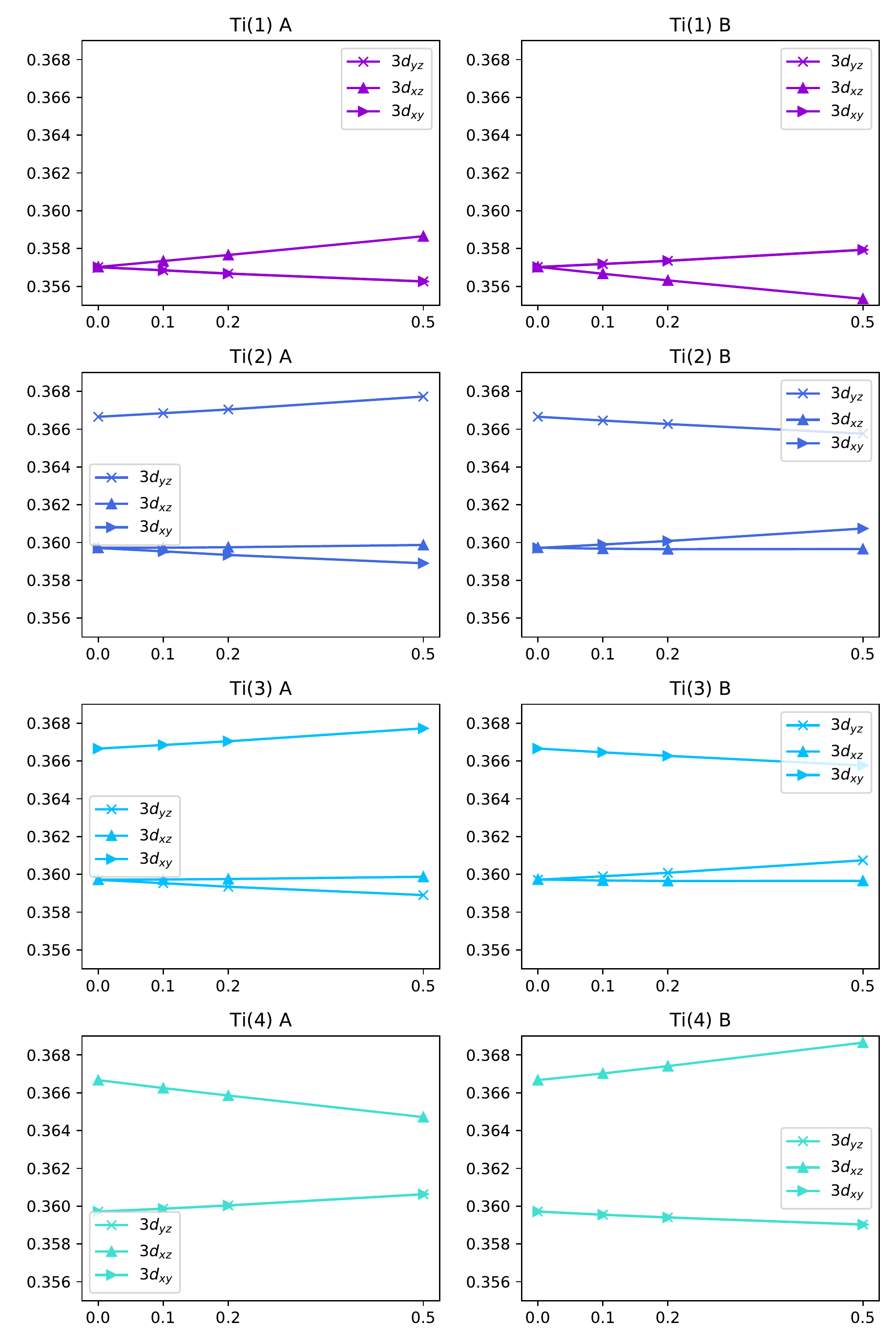}
  \caption{\textbf{Orbital densities as a functions of the relative phase shift.}
  The densities of projected Ti $4p$ (left) and Ti $3d$ (right) orbitals are shown for the eight distinct Ti atoms in the monoclinic unit cell, as a function of $\delta$. Coordinates for Ti atoms in layers A and B are given in  Table~\ref{tab:coordinates} (note that the parameter $A$ in that table changes sign depending on the layer), while orbitals are denoted with respect to the local coordinate axes.}
     \label{SI_occs}
 \end{figure*}

\end{document}